\PassOptionsToPackage{table}{xcolor}
\documentclass[sigconf,nonacm,9pt]{acmart}
\AtBeginDocument{%
	}

\usepackage{tikz}
\usepackage{amsmath}

\usepackage{mathrsfs, amsmath,amsfonts}
\usepackage{graphicx}
\usepackage{textcomp}
\usepackage{mdframed}

\usepackage{lipsum}

\usepackage[utf8]{inputenc}
\usepackage[T1]{fontenc}

\usepackage[abbreviations]{foreign}
\usepackage[table]{xcolor}
\usepackage{algorithm}
\usepackage{algpseudocode}
\usepackage[most]{tcolorbox}

\definecolor{hostcolor}{RGB}{220, 230, 255}  
\definecolor{pimcolor}{RGB}{230, 255, 230}   
\definecolor{transfercolor}{RGB}{255, 230, 230} 

\definecolor{codegreen}{RGB}{28,172,0} 
\definecolor{codeblue}{RGB}{51,102,255} 
\definecolor{codered}{RGB}{255,51,51} 

\newtcolorbox{greenline}[1][]{%
	colback=codegreen!10, 
	colframe=codegreen!50, 
	arc=2pt, 
	boxrule=0.5pt, 
	left=0pt, right=2pt, top=2pt, bottom=2pt, 
	#1 
}

\newtcolorbox{blueline}[1][]{%
	colback=codeblue!10, 
	colframe=codeblue!50, 
	arc=2pt, 
	boxrule=0.5pt, 
	left=0pt, right=2pt, top=2pt, bottom=2pt, 
	#1 
}

\usepackage[labelfont=bf,labelsep=period]{caption}
\usepackage[nodayofweek]{datetime}
\usepackage[inline]{enumitem}
\usepackage{float}
\usepackage{geometry}
\usepackage{graphicx}
\usepackage[utf8]{inputenc}
\usepackage{times}
\usepackage{url}

\usepackage{xspace}
\usepackage{pifont}

\setlist{itemsep=4pt, topsep=5pt}

\floatstyle{ruled}
\newfloat{algo}{htbp}{algo}
\floatname{algo}{Algorithm}

\def \ifempty#1{\def\temp{#1} \ifx\temp\empty }

\usepackage{textcomp}
\usepackage{booktabs} 
\usepackage{multirow}
\usepackage{tabularx}
\usepackage{makecell}
\usepackage{float}
\usepackage[export]{adjustbox}
\usepackage{eso-pic}

\usepackage{calc}

\usepackage{needspace}

\usepackage{code/codestyles}
\AtBeginEnvironment{algorithmic}{\linespread{1.1}\selectfont}

\renewcommand{\arraystretch}{1}

\newcolumntype{R}{>{\raggedleft\arraybackslash}X}
\newcolumntype{P}[1]{>{\raggedleft\arraybackslash}p{#1}}

\definecolor{prioritycolor}{HTML}{969bce}
\definecolor{darkgray}{HTML}{262626}

\usepackage{fontawesome5}
\newboolean{showcomments}
\setboolean{showcomments}{true}
\ifthenelse{\boolean{showcomments}}
{ \newcommand{\mynote}[3]{
		\fbox{\bfseries\sffamily\scriptsize#1}
		{\small$\blacktriangleright$\textsf{\emph{\color{#3}{#2}}}$\blacktriangleleft$}}}
{ \newcommand{\mynote}[3]{}}

\newcounter{numobserv} 
\definecolor{beaublue}{rgb}{0.88, 0.93, 0.93}

\usepackage[most]{tcolorbox}
\colorlet{shadecolor}{beaublue}
\tcbset{
	colback=beaublue,
	boxrule=0.5pt,
	boxsep=0pt,
	left=4pt,
	right=4pt,
	top=4pt,
	bottom=4pt,
	sharp corners,
	colframe=gray!70,
}
\newcommand{\observ}[1]{
	\addtocounter{numobserv}{1}
	\begin{tcolorbox}	
		\textit{\textbf{Take-away\,\thenumobserv\,:} #1 }	
	\end{tcolorbox}
}

\lstnewenvironment{listing}[1][]
{\lstset{#1}\pagebreak[0]}{\pagebreak[0]}

\usepackage{titlesec}
\renewcommand{\thesubsubsection}{\arabic{subsubsection}.}
\titleformat{\subsubsection}[runin]
{\normalfont\bfseries\itshape}
{\thesubsubsection}{0.5em}{}[\hspace{0.5em}\\~\\]

\newcommand{\code}[1]{\texttt{\small #1}\xspace}

\newcommand{\subpoint}[1]{\smallskip\noindent\textbf{#1}\xspace}

\newcommand{\tbb}{\texttt{tBB}\xspace}
\newcommand{\tob}{\texttt{tOB}\xspace}

\usepackage{hyperref}
\usepackage{xspace}
\newcommand{\sys}{\textsc{pim-cache}\xspace}

\newcommand{\copyrighttext}{ \scriptsize \textcopyright 2026 ACM.               
	Personal use of this material is permitted.                                 
	Permission from ACM must be obtained for all other uses,                   
	in any current or future media, including reprinting/republishing this      
	material for advertising or promotional purposes, creating new collective   
	works, for resale or redistribution to servers or                           
	lists, or reuse of any copyrighted component of this work in other works.   
	This is the author’s version of the work. The final version is published in the proceedings of the 27th ACM/IFIP International Middleware Conference. 
}

\setcopyright{none}
\begin{document}
	
	\title{PIM-CACHE: High-Efficiency Content-Aware Copy for Processing-In-Memory}	

	\author{Peterson Yuhala}
	\affiliation{%
		\institution{University of Neuchâtel}
		\city{Neuchâtel}
		\country{Switzerland}
	}
	\email{peterson.yuhala@unine.ch}
	
	\author{Mpoki Mwaisela}
	\affiliation{%
		\institution{University of Neuchâtel}
		\city{Neuchâtel}
		\country{Switzerland}	
	}
	\email{mpoki.mwaisela@unine.ch}
	
	\author{Pascal Felber}
	\affiliation{%
		\institution{University of Neuchâtel}
		\city{Neuchâtel}
		\country{Switzerland}
	}
	\email{pascal.felber@unine.ch}\author{Valerio Schiavoni}
	\affiliation{%
		\institution{University of Neuchâtel}
		\city{Neuchâtel}
		\country{Switzerland}	
	}
	\email{valerio.schiavoni@unine.ch}

	\date{\today}
	

	\newcommand{\copyrightnotice}{\begin{tikzpicture}[remember picture,overlay]       
			\node[anchor=south,yshift=2pt,fill=yellow!20] at (current page.south) {\fbox{\parbox{\dimexpr\textwidth-\fboxsep-\fboxrule\relax}{\copyrighttext}}};
		\end{tikzpicture}
	}

\begin{abstract}
Processing-in-memory (PIM) architectures bring computation closer to data, reducing the processor-memory transfer bottleneck in traditional processor-centric designs. 
Novel hardware solutions, such as UPMEM’s in-memory processing technology, achieve this by integrating low-power \emph{DRAM processing units} (DPUs) into memory DIMMs, enabling massive parallelism and improved memory bandwidth.
However, paradoxically, these PIM architectures introduce mandatory coarse-grained data transfers between host DRAM and DPUs, which often become the new bottleneck.
We present \sys, a lightweight data staging layer that dynamically eliminates redundant data transfers to PIM DPUs by exploiting workload similarity, achieving \emph{content-aware copy} (CAC).
We evaluate \sys on both synthetic workloads and real-world genome datasets, demonstrating its effectiveness in reducing PIM data transfer overhead.

\end{abstract}

	\maketitle
	\copyrightnotice
	

\section{Introduction}
\label{sec:intro}
Many modern workloads, such as genome analysis, graph processing, big data analytics, and image processing are data-intensive, meaning their performance is primarily limited by memory bandwidth.
This problem, known as the \emph{von Neumann bottleneck}, arises when the CPU processes data faster than it can be supplied from memory via the memory channel~\cite{rowclone23}.
\emph{Processing-in-memory} (PIM) is a paradigm shift from this traditional processor-centric design towards a more memory-centric design, where data can be processed where it resides~\cite{mutlu2019processing}.
Our work focuses on UPMEM's PIM architecture~\cite{upmem-pim} (henceforth UPMEM-PIM), the first PIM system to be commercialized~\cite{gomez22}. 

UPMEM-PIM integrates low-power RISC processors, known as \emph{DRAM processing units} (DPUs), into standard DRAM DIMM modules, with each DPU managing a 64MB chunk of memory, called \emph{main RAM} (MRAM).
This architecture provides extensive parallelism and achieves an aggregate memory bandwidth of up to $2.56$TB/s~\cite{upmem-pim}.
Several recent works~\cite{gomez22, upmem-pim,nider21, saransh21, mpoki24} have evaluated UPMEM-PIM on real-world workloads, showcasing the large performance improvements that can be obtained with this PIM architecture.
\begin{figure}[!ht]
	\centering	
	\includegraphics[scale=0.7]{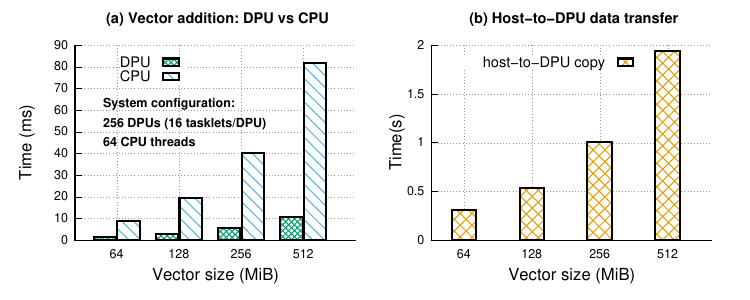}
	\caption{Total execution times for vector addition on two vectors of varying sizes, along with the total host-to-DPU copy time. Note: The copy time includes the total time required to transfer both input buffers of the corresponding size from host DRAM to DPU MRAM.}
	\label{fig:naive-va}
\end{figure}

Yet, ironically, the UPMEM-PIM architecture suffers from a data transfer bottleneck stemming from its architectural design, which maintains separate physical address spaces for conventional DRAM and PIM memory, \ie MRAM. 
Consequently, all data processed by UPMEM's PIM DPUs must be explicitly copied from DRAM to MRAM, incurring significant overhead. 
For example, \autoref{fig:naive-va} (a) illustrates the execution times for a vector addition benchmark on two input vectors of varying sizes, comparing performance on a UPMEM-PIM-enabled system with 256 DPUs (@400Hz) to a CPU-based approach (CPUs @2.1GHz).
The extensive parallelism of the PIM-based system allows for over a $7\times$ speedup with respect to the CPU-based approach.
However, \autoref{fig:naive-va} (b) highlights the significant cost of copying both input vectors to the DPUs, which completely dwarfs the performance gains achieved with the PIM-based system.

Past work~\cite{pim-mmu, friesel23} has identified the primary cause of this copy overhead as the inability for host to DPU data transfers to fully harness memory-level parallelism (MLP) via fine-grained interleaving across multiple DRAM chips and banks~\cite{zhang-interleaving}.
This is because a DPU has access to a single DRAM bank, and so data destined for the DPU must be transposed to ensure it ends up in a single bank, rather than interleaved across several banks.
Some prior studies like~\cite{gomez22, saransh21, upmem-pim} do extensive comparisons of PIM against CPU baselines but tend to overlook the CPU-DPU copy overhead when considering the speedup achieved with DPUs.
While such approaches may be justified in specific scenarios, they are generally unrealistic~\cite{friesel23}, as kernel-only benchmarks cannot accurately reflect real-world application behaviour~\cite{bailey91, heiser-crimes}.
Meanwhile, other studies~\cite{gilbert24,lee2024, mpoki24, friesel23, hyun24} have acknowledged this copy overhead and the scalability limitations it introduces to UPMEM's PIM architecture.

We observe that several PIM-friendly workloads, including machine learning training~\cite{gomez-ml-pim,roy2021pim}, graph analytics~\cite{pimpam, common-graph23}, genome analysis~\cite{alser2020accelerating, ben2024dart} among others, exhibit both temporal and spatial redundancy: similar data blocks are transferred across jobs or within large datasets. 
However, the current UPMEM-PIM CPU-DPU data copy approach is \emph{content-oblivious}, transferring bytes regardless of whether identical blocks already reside in PIM memory, thereby missing an opportunity to reduce overhead.
Our work bridges this gap by co-designing content-awareness with the memory-copy semantics of current PIM architectures like UPMEM. 

We propose \sys, a systematic \emph{content-aware copy} (CAC) approach with high efficiency for UPMEM-PIM. 
\sys leverages inline deduplication and fast byte-oriented compression~\cite{vbyte} to mitigate the expensive data copy overhead inherent in PIM architectures like UPMEM's.
Specifically, our CAC approach identifies repeating data patterns using well-known fingerprinting techniques, deduplicating the data before transferring it to DPU memory.
Further, our system retains data blocks in DPU memory for future use, thus preventing redundant data transfers for subsequent datasets. 
Additionally, we employ fast byte-oriented compression~\cite{vbyte} to further minimize the data copy overhead to DPUs.
Our CAC approach offers a practical solution for PIM programmers, obviating the need to implement ad-hoc data reduction techniques for workloads that are likely to contain repetitive data patterns.
We integrate \sys as a lightweight data staging layer in UPMEM's SDK, which we plan to release as open-source. 

In summary, this paper provides the following contributions: 
\begin{itemize}
\item Content-aware copy approach which exploits data similarity to mitigate CPU-DPU data transfer overhead in UPMEM DPUs; we provide this tool as open-source to the community
\item Support for lightweight byte-oriented integer compression for UPMEM DPUs
\item Extensive evaluations of these tools on synthetic and real-world datasets
\end{itemize}

\subpoint{Roadmap.}
The rest of this paper is organized as follows.
\S\ref{sec:background} presents important background concepts and the motivation behind our work. \S\ref{sec:arch} presents the system design behind our content-aware copy approach, and \S\ref{sec:implem} discusses the corresponding implementation details.
We evaluate our system in \S\ref{sec:eval}, discuss related work in \S\ref{sec:rw}, and finally conclude our paper in \S\ref{sec:conclusion}.

\section{Background}
\label{sec:background}
We first describe UPMEM's processing-in-memory infrastructure (\S\ref{sec:pim}), deduplication and compression (\S\ref{sec:dedup-vbyte}), after which we discuss the motivation behind our work (\S\ref{sec:motivation}) as a motivational application which drives our work.

\begin{table*}[!t]
	\caption{Redundancy characteristics in PIM workloads.}
	\label{tab:workload-redundancy}
	\small
	\setlength{\tabcolsep}{4pt}
	\renewcommand{\arraystretch}{1.15}
	\rowcolors{1}{gray!0}{gray!10}
	\begin{tabularx}{\textwidth}{@{}p{0.18\textwidth} p{0.15\textwidth} X@{}}
		\toprule
		\textbf{PIM Workload} & \textbf{Redundancy Type} & \textbf{Justification} \\
		\midrule
		Machine learning training & Temporal & ML training loops repeatedly access identical data batches across epochs~\cite{jeon19,tgopt23}.\\
		Comparative genomics & Spatial \& Temporal & Specimens belonging to the same species share $>$99\% of DNA; reference genomes are reused across jobs and comparative analyses~\cite{ben2024dart, dunham2013contemporary}. \\
		Iterative graph analytics & Temporal & Iterative algorithms like PageRank, BFS, \etc, process the same graph structure across multiple rounds and epochs~\cite{common-graph23, tegra21,  basak21}. \\		
		Log analytics & Spatial \& Temporal & Repeated headers, status codes, and error messages appear across millions of log entries~\cite{hpc-log-analytics}. \\		
		\bottomrule
	\end{tabularx}
	\vspace{-3mm}
\end{table*}

\subsection{Processing-in-memory}
\label{sec:pim}
\begin{figure}[!t]
	\centering	
	\includegraphics[scale=0.5]{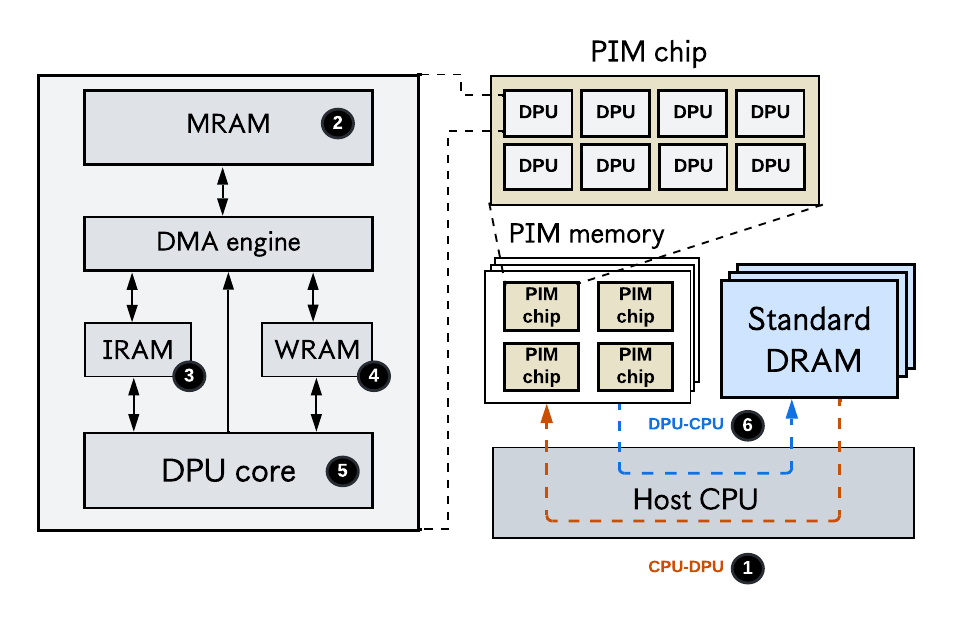}
	\caption{\vspace{-10pt}Architecture of a UPMEM-PIM enabled system.}
	\label{fig:pim-arch}
\end{figure}
Processing-in-memory (PIM) (or near-memory processing), is a computing  paradigm which augments memory with computation capability, mitigating the data movement bottleneck arising from limited memory bandwidth in traditional CPU-centric processing paradigms.
Several PIM proposals exist~\cite{agrawal17, ahn15, gao16}. 
UPMEM-PIM~\cite{upmem-pim} is the first general-purpose PIM hardware to become commercially available.
In the following, we details the UPMEM-PIM architecture and provide relevant details on UPMEM-PIM programming paradigms.

\subpoint{UPMEM-PIM architecture.}
\autoref{fig:pim-arch} shows the architecture of a UPMEM-PIM-enabled system.
It comprises a host CPU, main memory (\ie DRAM), and PIM-enabled memory equipped with UPMEM-PIM modules.
These are are standard DDR4-2400 DIMMs comprising 2 DRAM ranks, with 8 PIM chips per rank.
Each chip comprises 8 general-purpose processing cores, called \emph{DRAM processing units} (DPUs), thus 64 DPUs per rank and 128 DPUs per PIM module.

Data to be processed by PIM modules must be transferred~\ding{202} from main memory to DPU \emph{main RAM}~\ding{203} (MRAM), a 64MB DRAM memory bank shared with the host CPU.
DPUs have no memory management unit (MMU), so programmers must derive the physical memory addresses in MRAM of data to be copied to the DPU.
Each DPU contains $88$KB of SRAM which is split into 24KB instruction memory called instruction RAM (IRAM)~\ding{204}, and 64KB scratchpad memory called working RAM (WRAM)~\ding{205}.
UPMEM's SDK provides instructions to move data between WRAM and MRAM via DMA.
The DPU core~\ding{206} is a 14-stage interleaved pipeline processor which uses $24$ hardware threads to achieve scaling.
Results from DPU processing can equally be transferred from PIM memory to main memory~\ding{207}.

\subpoint{Why explicit host-DPU data transfers?} 
UPMEM-PIM uses separate address spaces for the host and DPUs because the DPUs do not support hardware cache coherence with the host CPU. Consequently, the DPU architecture enforces exclusive access to the MRAM bank, either by the host CPU or by a DPU at any given time, thereby preventing memory access conflicts.
Implementing a unified address space would require complex hardware modifications to the memory controller to properly arbitrate host CPU and PIM DPU memory accesses~\cite{nider20,lee2024, pim-mmu}. 
As a result of this address space separation, data must be explicitly copied from the CPU's address space (standard cache-coherent DRAM) to the DPU's MRAM.
All inter-DPU communication must also be routed through the host CPU.

\subpoint{Why are the host-DPU data transfers costly?}
In standard DRAM architectures, memory is interleaved~\cite{zhang-interleaving}, meaning that consecutive memory addresses are distributed across multiple memory banks (and chips) rather than residing sequentially within a single bank. 
As a result, when the CPU writes a buffer, \eg 8-bytes of size, to a DIMM, the data is spread across 8 different physical DRAM chips in a DRAM rank (each chip taking 1 byte of data), so as to optimize performance via memory-level parallelism (MLP).	
However, in PIM designs such as UPMEM, this interleaving could lead to incorrect computations, as DPUs have access only to their local MRAM banks. 
Instead, the buffer for DPU computation must be fully copied into the corresponding DPU's MRAM bank. 
This requirement necessitates data transposition during copy operations to ensure the data ends up in a single bank~\cite{pim-mmu,friesel23,nider21}, rather than in separate banks across several chips, thus preventing host to DPU transfers from fully harnessing MLP via fine-grained interleaving across channels, ranks, or banks. 
The absence of MLP via interleaving prevents the full DRAM/MRAM write bandwidth from being achieved; this, together with data transposition operations during host to DPU data transfers, explain the poor host to DPU data transfer performance. 


\subpoint{UPMEM-PIM programming model.}
A UPMEM-PIM program consists of two parts: (1) a host program which executes on the CPU, and (2) a PIM kernel which executes on the DPUs.
PIM kernels follow the single program multiple data (SIMD) programming model, and $24$ software threads called \emph{tasklets} map to the DPU's hardware threads allowing for extensive parallelism in UPMEM's PIM infrastructure.
The host program performs PIM orchestration operations: initializing DPUs, transferring data to them, and launching the DPUs.
At the time of this writing, host programs can be implemented in C, C++, Java, or Python, while the DPU code must be implemented in C.

\subsection{Deduplication and compression}
\label{sec:dedup-vbyte}
Deduplication and compression are both data reduction techniques that remove redundant content at different granularities~\cite{austere-cache, chunkstash, wu18}.

\subpoint{Deduplication.}
Deduplication splits data into multiple non-overlapping data units referred to as \emph{chunks} or \emph{blocks}, typically of size in the order of KiB.
Each data chunk is uniquely identified by a fingerprint (FP) which is computed using either a cryptographic or non-cryptographic hash function (\eg SHA-1, XXHash) over the chunk's content.
Chunks with identical fingerprints are treated as duplicates.
The deduplication system stores mappings of hashed chunks and their corresponding locations (\eg logical block address in storage, or memory offsets) in an indexed structure (\ie hash table), used for duplicate checking in subsequent chunks.

Deduplication can either be \emph{inline} (it happens before data is written or transferred), or \emph{post-process}, done after data is transferred or stored, typically on secondary storage.
Our work follows the inline approach, as it aims to reduce data before it is transferred to DPUs.

\subpoint{Byte-oriented integer compression.}
In contrast to deduplication, which operates on larger blocks of data, compression works at byte-level granularity for data reduction.
We focus on \emph{byte-oriented compression} for 32-bit unsigned integers.
This technique encodes integers such that the main data is stored in consecutive whole bytes, rather than a predefined number of bytes (\eg 4 bytes for 32-bit unsigned integers).
A widely used byte-oriented integer compression technique is \emph{variable byte encoding} (VByte)~\cite{vbyte} which efficiently encodes integers using a variable number of bytes: it splits each integer into 7-bit chunks, reserving the most significant bit as a continuation bit to indicate that the following byte continues the current integer.
As such, integers in the range $[2^{L-1}, 2^{7L})$ are encoded using $L$ bytes, with $L=1,2,\ldots$.
For example, the integer $64$ is encoded in a single byte (\textbf{0}1000000), while $128$ requires two bytes (\textbf{1}0000000 and \textbf{0}0000001).
Byte-oriented compression techniques are commonly used in search engines, database indexing, information retrieval~\cite{stepanov11, dean2009, vbyte}, and even genome analysis~\cite{brandon2009data}.

\section{Motivation}
\label{sec:motivation}

PIM DPUs are typically deployed as high-bandwidth coprocessors, where large datasets are partitioned and offloaded across jobs or iterations, for in-memory processing.
Common PIM-friendly workloads like machine learning (ML)~\cite{gomez-ml-pim,roy2021pim}, genome processing~\cite{alser2020accelerating}, log analysis~\cite{hpc-log-analytics}, \etc, exhibit two key forms of redundancy which can be exploited to mitigate data transfer overhead:
\begin{itemize}[nosep,leftmargin=*]
\item \textbf{Temporal redundancy}: identical data blocks reappear \emph{across transfers}, \eg same training batch for ML training, or a genome sequence copied after a reference genome has been copied to a DPU.

\item \textbf{Spatial redundancy}: duplicate data blocks exist \emph{within a single buffer}, \eg blocks of zeros in a sparse matrix.
\end{itemize} 
\autoref{tab:workload-redundancy} shows that these patterns are systematic, making them ideal targets for optimization. 
Yet, the current UPMEM-PIM memory-copy semantics remains \emph{content-oblivious}, forcing PIM programmers to implement ad hoc, workload-specific hacks to mitigate transfer overhead.

\sys bridges this gap by co-designing \emph{content-awareness} with PIM's memory-copy semantics, leveraging deduplication and compression techniques to eliminate redundant data transfers from the host CPU to DPUs.

\section{System design}
\label{sec:arch}
\begin{figure*}[!ht]
	\centering	
	\includegraphics[scale=0.7]{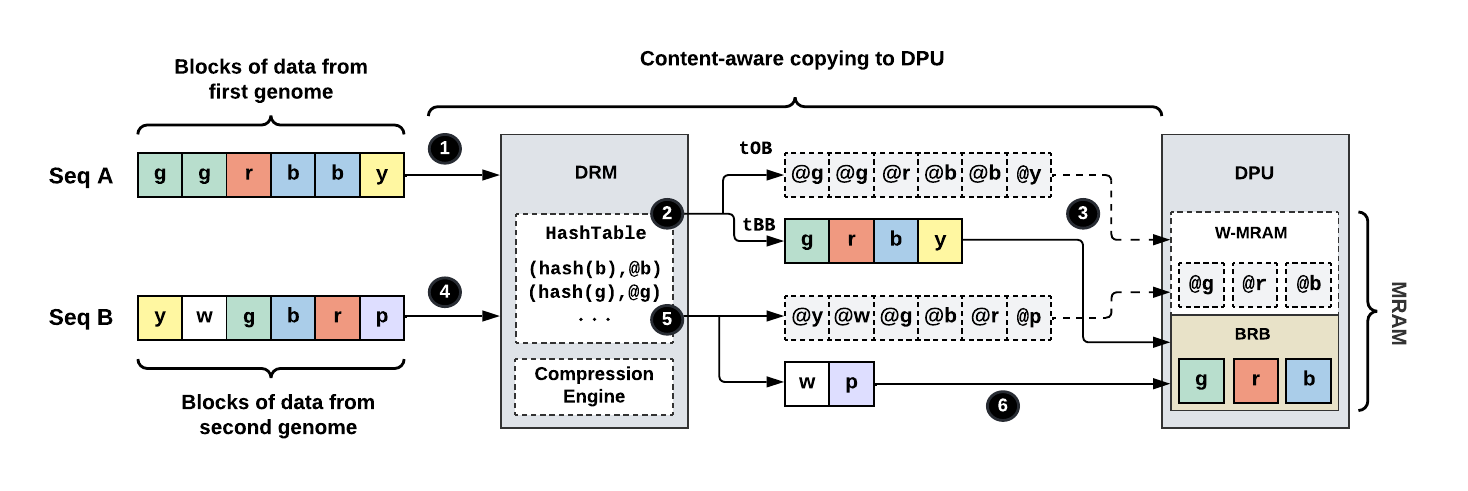}
	\caption{Content-aware copy design.	\vspace{-10pt}}
	\label{fig:arch}
\end{figure*}

We present here the design and inner workings of \sys.
We use a simple illustrative PIM-based application: comparative genome analysis on two genome sequences \(\mathsf{Seq~A}\) and  \(\mathsf{Seq~ B}\) represented as a sequence of fixed-sized blocks, \eg 1KiB.
The same ideas are applicable to a non-genomic workloads.
\autoref{fig:arch} illustrates the full workflow.
As depicted in the inputs: \(\mathsf{Seq~A}\) and \(\mathsf{Seq~B}\), we use specific colours: green (g), red (r), blue (b), yellow (y), white (w), or purple (p) to indicate blocks with identical content, \eg sets of identical sequences of DNA bases: \(\mathsf{GATTACGCA\ldots}\).
Given the high likelihood of both temporal and spatial redundancies in such workloads, our system optimizes CPU to DPU data transfers by avoiding redundant copy operations.

\subsection{Challenges}
Several challenges must be addressed to correctly integrate a content-aware copy approach into a PIM architecture like UPMEM.

\subpoint{Efficient data management across DPUs.}
The first challenge involves efficiently tracking and managing data across hundreds to thousands of different DPUs. 
Traditional deduplication approaches, \eg those designed for secondary storage, assume a single unified storage space.
In contrast, a PIM system like UPMEM distributes memory across thousands of independent DPUs.
To put this into perspective, a fully-equipped UPMEM PIM server contains 2560 DPUs at the time of this writing.
This architectural structure introduces several complexities involving efficient data splitting, deduplication and tracking of the state of each DPU's memory. 
This equally requires thinking about data replacement across the DPUs.

\subpoint{Data reconstruction on limited DPU memory.}
Deduplication splits data into chunks and stores only one physical copy of each unique block. 
However, when the data must be used, \ie processed in DPUs, the system needs to recreate the exact original byte sequence, in the original order, from the deduplicated metadata.
This is particularly challenging when dealing with PIM DPUs with relatively small memory (\ie $64$MB of MRAM and $64$KB of WRAM) when compared to secondary storage systems with terabytes of storage capacity. 

The rest of this section discusses how these challenges are addressed to achieve a content-aware copy design in UPMEM PIM.

\subsection{Content-aware copying to PIM DPUs}
At the core of our solution is the \textbf{content-aware copy (CAC)} approach.
It leverages prior knowledge of data content, \ie data blocks, to decide whether to transfer them or not to DPUs.
We partition each DPU's MRAM into two regions: a \emph{block retention buffer} (BRB) which retains distinct data blocks copied into MRAM, and a smaller "working" MRAM (W-MRAM) region reserved for metadata and any in-place computations.
Our CAC approach consists of three main steps (\autoref{fig:arch}): (\ding{202})~\emph{data chunking for DPUs}, (\ding{203})~data deduplication and (\ding{204})~\emph{content-aware data transfer}.
Steps \ding{205}, \ding{206}, and \ding{207} are analogous to \ding{202}, \ding{203}, and \ding{204} respectively, performed on the subsequent input \(\mathsf{Seq~B}\) transferred to the DPU.
We include these additional steps to provide a clearer view of the system's inner workings with multiple inputs.

Incoming data streams are first logically partitioned into equal chunks for DPUs, with each chunk uniquely identified by its start and stop offset in the data stream, \eg buffer.
These logical partitions are then fed to a \emph{data reduction module} (DRM) (\ding{202}), which fingerprints fixed-size data blocks in the data partition, \eg 512B or 1KiB blocks, using a fast hashing algorithm (XXHash~\cite{xxhash} in our case).
These fingerprints are used as keys to query a per-DPU hash table (\ding{203}) managed by the host application, which maps block fingerprints to their addresses (or offsets) in the corresponding DPU's BRB; each (per-DPU) hash table essentially tracks the state of the corresponding DPU's BRB (\ie which blocks it contains).
As such, a lookup miss in the hash table indicates a block is absent in the DPU's BRB; the block is appended to a temporary block buffer (\autoref{fig:arch} \tbb), while its offset (\ie destination address in the BRB/MRAM) is appended to a temporary offset buffer (\autoref{fig:arch} \tob).
A lookup hit, however, only records the block's offset in the \tob, effectively replacing a costly data transfer (\eg 1KiB) with a more lightweight metadata copy (\ie 32-bit integer offset).

Following our illustrative example, the first occurrence of block \texttt{g} from  \(\mathsf{Seq~A}\) results in a lookup miss in the hash table. 
Its offset \textbf{\texttt{$@$}}g is thus computed and the key-value pair \texttt{(hash(g), $@$g)} is inserted into the hash table. 
Block \texttt{g} is then appended to the \tbb and its offset \textbf{\texttt{$@$}}g  appended to \tob.

However, the second occurrence of block \texttt{g} from \(\mathsf{Seq~A}\) indicates a spatial redundancy (duplicates in the same buffer), and results in a lookup hit, so only its offset \textbf{\texttt{$@$}}g is added \tob. 
Similarly, when a subsequent data chunk, \eg sequence \(\mathsf{Seq~B}\) is fed to the DRM (\ding{205}), the hashing and lookup operations for blocks \texttt{y}, \texttt{g}, \texttt{b}, and \texttt{r}~(\ding{206}) will represent a temporal redundancy (duplicates across several buffers) and result in lookup hits, tracking only their BRB offsets in the \tob.   
At the end of DRM operation for each buffer, the \tbb (distinct data blocks) and \tob (all block offsets) are copied to the DPU's BRB and W-MRAM respectively~(\ding{204}).

\subpoint{Data reconstruction on DPUs.}
After deduplicating data destined for DPUs, we reduce both transfer costs and MRAM footprint by avoiding duplicate block copies. 
However, this introduces a new challenge: how to correctly reconstruct the original data in the DPU?
This is especially difficult considering UPMEM's strict memory constraints: each DPU has only $64$MB of MRAM and $64$KB of WRAM.

We address this by avoiding any physical reconstruction of the full buffer inside the DPU.
Instead, each DPU performs a \emph{logical reconstruction} of the original buffer using the ordered list of offsets (\ie block addresses in MRAM/BRB) in the \tob.
To do this, the PIM kernel iteratively fetches blocks of data using their addresses/offsets from \tob into limited WRAM memory, processes them, and writes the results back to W-MRAM.
Because the offsets in the \tob preserve the original ordering, the computation proceeds exactly as if the full input buffer had been reconstructed. 
For example, in a vector-add benchmark, the kernel sequentially loads the required blocks, performs addition operations on elements in each block, and writes the partial results to an output buffer in W-MRAM



\subpoint{BRB block replacement.}
When the BRB is full, \sys applies a \emph{global invalidation policy} in the corresponding DPU hash table (there is no actual ``data flushing''). 
We chose this approach over a more complex FIFO or LRU block-replacement strategy because it simplifies memory management and eliminates fragmentation, guaranteeing contiguous free space in a BRB.
For workloads exhibiting high temporal or spatial redundancy, and a sufficiently large BRB (\eg $90\%$ of MRAM), BRB invalidations are infrequent, effectively balancing memory overhead with the benefits of data reuse.

\subpoint{Compression.}
While deduplication excels at eliminating exact block-level duplicates across data transfers, it does not provide much benefit for non-redundant, one-time data transfers.
To broaden \sys's applicability, the DRM incorporates a lightweight \emph{compression engine} (CE) that employs variable-byte encoding (\S\ref{sec:dedup-vbyte}).
This is particularly effective for integer datasets with high compressibility under VByte encoding, \ie values in the ranges $[0, 2^7)$ and $[2^7, 2^{14})$.
Such data patterns are common in data sets used for genome analysis where nucleotide bases are encoded as small positive integers in this range, \ie A = 0, C = 1, G = 2, T = 3~\cite{schbath2012mapping,brandon2009data} and in information retrieval~\cite{stepanov11,im-pir}.
The CE partitions input buffers across host CPU threads for parallel compression, then transfers the compressed payload to DPUs.
A corresponding decompression engine on the DPU side leverages the large parallelism from by DPU tasklets to parallelize VByte decompression, offering significant data transfer improvements overall (\S\ref{eval:compression}).

\subpoint{Implementation.}
We use XXhash64~\cite{xxhash}, a fast hashing algorithm with high collision-resistance, for block fingerprinting. 
The core of our content-aware copy approach, the DRM deduplication logic, comprises $\approx$ 500 lines of ANSI-C code (LoC). 
The byte-oriented integer compression/decompression logic comprises $\approx$ 300 LoC.
%
\section{Implementation}
\label{sec:implem}

\subpoint{Block fingerprinting.}
Our implementation uses XXhash64~\cite{xxhash}, a fast hashing algorithm for block fingerprinting (see \S\ref{eval:block-hashing}). 
XXhash64 generates 64-bit fingerprints, thus providing significantly higher resistance to hash collisions when compared to 32-bit fingerprinting algorithms.

We leverage CPU-level multi-threading to speed up computation of block fingerprints and hash table lookups and insertions.
Our multi-threaded implementation partitions blocks of data across threads in such a way that ensures only one thread accesses a DPU's hash table during lookup and insertion operations by the DRM.
This prevents any race conditions due to multiple threads updating a DPU's data, and equally guarantees that the order of block addresses/offsets in the temporary offset buffers matches the order of the blocks in the portion of the input buffer destined for each DPU.
In our prototype implementation, the DRM deduplication logic comprises $\approx$ 500 lines of code (LoC), while compression/decompression logic comprises $\approx$ 500 LoC.

\section{Evaluation}
\label{sec:eval}
This section evaluates the performance of the data reduction strategies proposed.
It answers the following questions:

\begin{itemize}[]
	\item[\textbf{Q1}:] What is the computational overhead introduced by our CAC approach? (\S\ref{eval:drm})
	\item[\textbf{Q2}:] How effective is CAC in reducing CPU-DPU data transfer overhead in synthetic and real-world workloads? (\S\ref{eval:dedup} )
	\item[\textbf{Q3}:] How effective is VByte compression in reducing CPU-DPU data transfer overhead? (\S\ref{eval:compression})
	\item[\textbf{Q4}:] What are the overall end-to-end performance benefits of CAC? (\S\ref{eval:e2e})
	\item[\textbf{Q5}:] What is the memory overhead introduced by CAC? (\S\ref{eval:mem})
\end{itemize}


\subsection{Experimental setup}
\subpoint{System configuration.}
The experimental machine is a dual-socket server equipped with two 16-core Intel(R) Xeon(R) Silver 4216 processors clocked at 2.10 GHz, and 251GB of DRAM.
The server comprises 4 UPMEM DDR4-2400 DIMMs, each containing 128 DPUs, for a total of 512 DPUs.
Each DPU operates at 400MHz and is equipped with 64MB of MRAM, resulting in a total of 32GB of MRAM.
The memory bandwidth between each DPU and its associated MRAM bank is 1GB/s, representing an aggregate DPU memory bandwidth of up to 512GB/s across the server.
The server runs Ubuntu 22.04.5 LTS.

\subpoint{Workloads.}
We evaluate our system on both synthetic and real-world workloads.
\begin{enumerate}
\item{\textbf{Synthetic}.}
We generate synthetic data to account for various degrees of data redundancy. 
The synthetic workloads are segmented arrays of randomly generated integers with varying degrees of controlled repetition/entropy, quantified as a repetition ratio $R$ = ratio of repeated values in each segment. 
We have $0\leq R\leq 1$, where $R=1$ depicts the highest degree of repetition (lowest entropy), while $R=0$ the least degree of repetition (highest entropy).


\item{\textbf{Genome data}.}
Genomic analysis involves processing large datasets that often contain repetitive patterns, making CAC a promising approach in this area.
We use several aligned genome sequences from the UCSC Genome Browser~\cite{ucsc,perez2025ucsc}, an online database used for bioinformatics.
Aligned genome sequences are often compared to a reference genome, \eg the human reference genome (GRCh38~\cite{grch38, schneider2017evaluation}) to determine genetic variations or similarities between the genomes.
We simulate a PIM-based comparative genomic analysis by copying these genome sequences into PIM memory in successive CPU-DPU transfers, illustrating how CAC tackles temporal redundancy. 


\end{enumerate}

Unless stated otherwise, the reported results are medians over 50 independent runs.

\subsection{DRM processing overhead}
\label{eval:drm}
\begin{figure*}[!t]
	\centering	
	\includegraphics[scale=1]{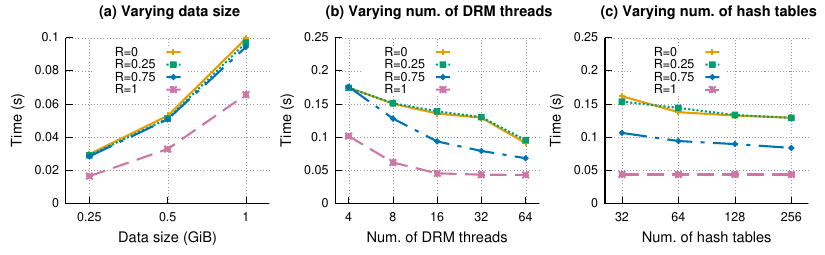}
	\caption{Overhead of DRM operations with varying number of DRM threads, hash table, and data sizes.}
	\label{fig:drm-overhead}
\end{figure*}
We begin by evaluating the cost of DRM operations on synthetic workloads while varying three parameters: data size, number of DRM threads, and number of DRM hash tables (each hash table tracks a DPU's BRB state).

\subpoint{Varying data size.}
\autoref{fig:drm-overhead} (a) depicts the overall cost of DRM operations as the input buffer size varies, while keeping the number of DRM threads fixed at $32$ and the number of hash tables (hence target DPUs) at $256$.
We observe that the overall cost of DRM operations increases with data size, which is a result of more data blocks and fingerprinting operations being performed per DRM thread.
Workloads with very high redundancy ($R=1$) exhibit lower overhead when compared to those with lower redundancy (\eg $R=0$).
This is primarily due to improved cache locality during block hashing operations, as repeating data patterns allow for more efficient memory access.
Additionally, the presence of redundant blocks accelerates hash table lookups when employing last-key optimization techniques (\S\ref{sec:arch}) which avoid redundant hash table queries when the same block fingerprint was previously queried in the hash table.

\subpoint{Varying number of DRM threads.}
\autoref{fig:drm-overhead} (b) illustrates the cost of DRM operations when varying the number of DRM threads, while keeping the input data size fixed at $1$GiB and the number of hash tables at $256$.
Overall, using a larger number of threads improves DRM operation speed because we have block fingerprinting as well as hash table lookups being done in parallel. 
We note that the DRM is designed such that input blocks and hash tables are partitioned across the threads, preventing multiple threads from accessing the same blocks or hash tables.
This minimizes the need for any synchronization operations among DRM threads.
Similarly, higher redundancy workloads exhibit better performance as explained previously. 

\subpoint{Varying number of DRM hash tables.}
\autoref{fig:drm-overhead} (c) illustrates the cost of DRM operations when varying the number of hash tables while keeping the input data size fixed at $1$GiB and the number of DRM threads at $32$.
The experimental results show that, for a given data size, the number of hash tables (thus target DPUs) managed by the DRM has minimal effect on overall DRM performance.
This suggests that block fingerprinting is the principal source of overhead, and not hash table lookup/insertion operations.
Nevertheless, we observe that larger redundancy in the dataset leads to better performance with respect to lower redundancy data sets, which, as previously explained, can be attributed to improved CPU cache locality during block fingerprinting operations.


\subsection{Impact of deduplication on copy overhead}
\label{eval:dedup}
\begin{figure*}[!t]
	\centering	
	\includegraphics[scale=1]{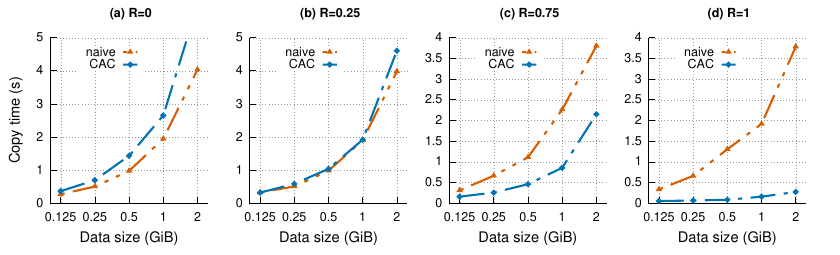}
	\caption{Host to DPU data transfer overhead with CAC and without CAC (naive) using synthetic workloads with varying degrees of spatial redundancy. The least redundant workload is $R=0$ while the most redundant is $R=1$. As spatial redundancy increases (from left to right), the benefits of CAC become more apparent. We use 256 DPUs.}
	\label{fig:copy-synthetic}
\end{figure*}
Now, we study the effect of CAC on data transfer overhead, first with synthetic workloads, and subsequently with real genome data.

\subpoint{Synthetic workloads.}
\autoref{fig:copy-synthetic} illustrates the variations in CPU-DPU transfer times across synthetic workloads with varying degrees of spatial redundancy, comparing our CAC approach to naive (content-agnostic) copying to DPUs.

For the least redundant workload $R=0$, our CAC approach exhibits poor performance.
This is expected as this workload contains no duplicates, hence a deduplication percentage\footnote{The deduplication percentage is computed as $dedup\_percentage = (1 - \frac{deduplicated~size}{original~size}) * 100$} of $0\%$ is achieved at the DRM.
The additional DRM operation overhead, as well as the cost of transferring block offsets, makes CAC about $1.3\times$ slower with $R=0$ compared to the naive copy approach.

However, as spatial redundancy increases in the workloads (moving from left to right in \autoref{fig:copy-synthetic}), CAC exhibits improved data transfer overhead, up to $14\times$ for the most redundant workloads like $R=1$.
This workload results in a deduplication percentage of over $98\%$, hence much fewer blocks are transferred to the DPUs, as opposed to the naive, content-agnostic approach which copies the full buffer regardless of data redundancies.
While this experiment considers spatial redundancy, we note that CAC's deduplication strategy goes beyond simply removing duplicate blocks in the current buffer being analyzed, as it also applies across multiple transfers, \ie temporal redundancy.
This means the results for high-redundancy workloads like $R=1$ or $R=0.75$ are equally representative of scenarios where the DPU already contains similar data in its BRB from prior data transfers, even if the current buffer being transferred shows little spatial redundancy.
For this reason, even in situations with spatial redundancy (\ie $R=0$), our CAC approach can still provide long-term benefits provided there temporal redundancies between the current and subsequent datasets. 
In the following, we illustrate this using real genome data.

\observ{
A content-aware copy approach optimizes data transfers when a buffer contains high spatial redundancy, reducing both data transfer time and memory usage in PIM DPUs.
}

\subpoint{Genome data.}
\label{eval:genome}
\begin{figure}[!t]
	\centering	
	\includegraphics[width=1\linewidth]{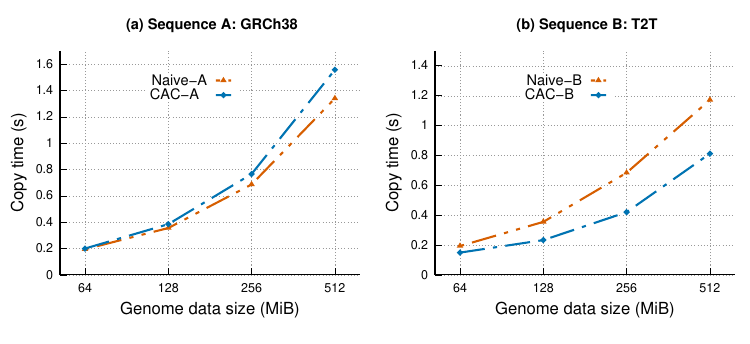}
	\caption{Data transfer times for genome sequences transferred successively to 256 DPUs with and without CAC. Sequence A little spatial redundancy, leading to relatively poor performance with respect to the naive copy approach. However, there are large temporal redundancies between sequence B and A, leading to improved data transfer performance for B.}
	\label{fig:genome-copy}
\end{figure}
In this section, we leverage real-world genome datasets to test our CAC approach.
We copy varying sizes of chunks of genome sequences to PIM DPUs: first the human reference genome (GRCh38) and then the T2T assembly of the human genome~\cite{aganezov2022complete}.
Both genomes are in FASTA format~\cite{fasta} with their headers removed (so contain only the nucleotide base sequences).
Performing comparative analysis of these two genome sequences is common in genomics~\cite{aganezov2022complete}.
We compare the time of doing these CPU-DPU data transfer operations using CAC against a naive copy approach.
\autoref{fig:genome-copy} illustrates the results.

First, we observe a performance loss of up to $1.16\times$ with our CAC approach when copying sequence A (GRCh38) to the DPUs (\autoref{fig:genome-copy} (a)).
While genome sequences contain highly repetitive regions within the same sequence~\cite{treangen2012repetitive}, these regions aren't as large as our default block size of 1KiB.
This is confirmed by the analysis of sequence A, where the DRM reports $0\%$ (block-level) spatial redundancy for a genome sequence of size $512$MiB, and just $2$ duplicates for a $1$GiB sequence ($2^{20}$ 1KiB blocks).
Prior work on DNA deduplication corroborates this observation~\cite{cogo21}.
However, when considering multiple separate and aligned genome sequences, a significant amount of temporal redundancy is observed.
This is demonstrated in \autoref{fig:genome-copy} (b) where transferring sequence B (T2T) after transferring GRCh38 results in $40\%$ duplicates on average, as similar blocks had already been processed in the previous sequence.
This leads to a performance gain of up to $1.5\times$ when compared to the naive approach.
Although some performance is lost via DRM operation and copying additional block offsets, the overall impact is a net improvement when considering all the data transfer operations.
In workloads such as ML model training where the same data set is used across hundreds of epochs (high temporal redundancy), \sys's content-aware copy will prevent any redundant copying.


\observ{
Even when spatial redundancy is low, a content-aware copy approach effectively optimizes data transfer when there are temporal redundancies across workloads, \eg in genome analaysis and ML model training across epochs.
}

\subsection{Effect of compression on data transfer overhead}
\label{eval:compression}
\begin{figure}[!t]
	\centering	
	\includegraphics[width=1\linewidth]{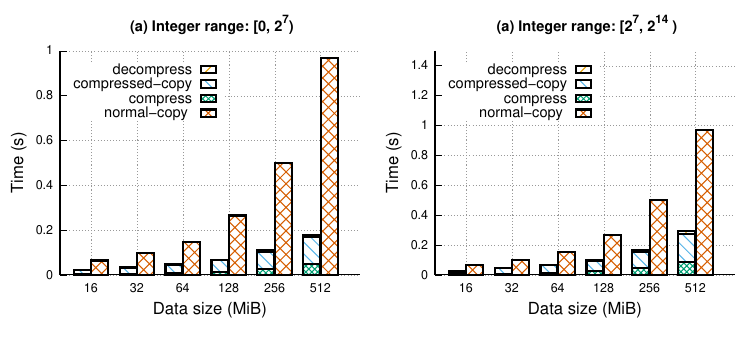}
	\caption{Effect of compression on data transfer overhead. Integers in $[0, 2^7)$ are represented using a single byte with VByte compression, while those in $[2^7, 2^{14})$ use two bytes.}
	\label{fig:compression}
\end{figure}
Here we evaluate the cost of transferring VByte-compressed arrays of varying sizes to 256 DPUs and compare these to the cost of transferring the uncompressed arrays.
We focused on integers in the ranges $[0, 2^7)$ (Group 1) and $[2^7, 2^{14})$ (Group 2), because they achieve the largest compression ratios, $4$ and $2$ respectively.\footnote{The compression ratio is computed as $ratio = \frac{original~size}{compressed~size}$} 
We used $16$ compression threads on the host, and $16$ tasklets per DPU for decompression.
The results with these parameters show that VByte compression reduces the overhead of CPU-DPU data transfers by up to $5.4\times$ for Group 1 and $3.3\times$ for Group 2, including both compression and decompression times.
While we expect these results to vary for different numbers of compression and decompression threads and data sizes, our results demonstrate the potential improvements in data transfer efficiency with such compression strategies.


\subsection{Choice of block size and fingerprinting algorithm}
\label{eval:block-hashing}
\begin{figure}[!t]
	\centering	
	\includegraphics[width=1\linewidth]{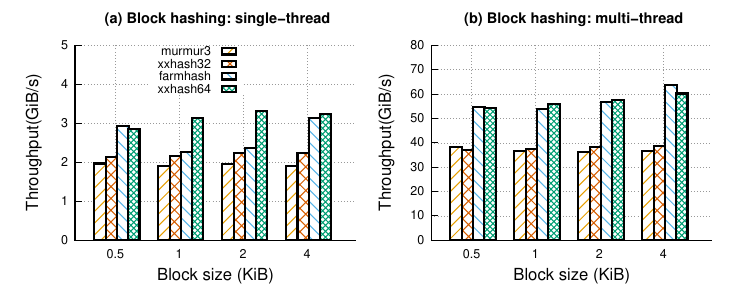}
	\caption{Fingerprinting throughput with different hashing algorithms}
	\label{fig:block-hashing}
\end{figure}
Here, we aim to understand how variations in block size may affect fingerprinting efficiency. 
This evaluation helps determine the block sizes for optimal deduplication performance.
We evaluated the cost of performing block fingerprinting on a $1$GiB input buffer using four well-known fingerprinting algorithms: Murmur3~\cite{murmur}, XXHash (32-bit and 64-bit)~\cite{xxhash}, and Google's FarmHash~\cite{farmhash}, while varying the block size.
First, we computed fingerprints for all blocks using a single thread, and then we split the blocks across $16$ threads which computed block fingerprints in parallel.\footnote{We note that we aren't using multi-threaded versions of the fingerprinting algorithms, but splitting the input blocks across multiple threads which use the single-threaded fingerprinting algorithms.}
\autoref{fig:block-hashing} depicts the results obtained.
The general observation is that the throughput of the fingerprinting process (for both single and multi-threaded versions) does not vary much for the block sizes chosen, with a slight increase in throughput for $4$KiB blocks when using FarmHash and XXHash64. 
While we could choose $4$KiB blocks as the default block size for DRM deduplication, this has two drawbacks: (1) it increases the granularity of the deduplication process, reducing the likelihood of duplicate blocks in the same dataset, and (2) as of this writing, DPU WRAM is limited at $64$KiB, meaning there is a higher chance of memory shortages when tasklets allocate large block sizes in WRAM.
This issue is further exacerbated by the fact that as of this writing, UPMEM's SDK only allows for WRAM allocations, but no deallocations; the only way being to free the entire heap (for all tasklets) via \code{mem\_reset()}~\cite{upmem-mem-man}.
Very small blocks sizes like $512$ bytes mitigate this, while improving deduplication granularity, but also introduce metadata overhead, \ie larger indexing tables. 
As a result, we opted for $1KiB$ as the default DRM block size, which is a good compromise.

\subsection{End-to-end performance evaluation}
\label{eval:e2e}
\begin{figure*}[h]
	\centering	
	\includegraphics[width=1\linewidth]{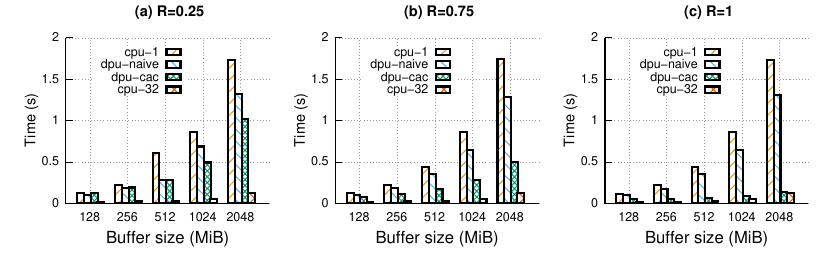}
	\caption{End-to-end performance evaluation of \sys with a vector addition workload. The buffer size accounts for all the vectors copied and processed, while the time comprises both data transfer and processing times for the PIM-based workloads.}
	\label{fig:e2e-eval}
\end{figure*}
To assess the overall benefits of using CAC, we executed a vector addition workload on buffers with varying degrees of redundancy under four configurations: CPU-based execution using a single thread (\texttt{cpu-1}), CPU-based execution using 32 threads (\texttt{cpu-32}), and DPU execution with (\texttt{dpu-cac}) or without CAC (\texttt{dpu-naive}).
The data size accounts for both vectors copied and processed. 
This experiment provides a full end-to-end assessment of our approach, as it covers both deduplication, copying, and processing on the DPUs, and compares this to a CPU-based approach. 
\autoref{fig:e2e-eval} outlines the results obtained. 

As discussed previously, the cost of copying data to DPUs dominates overall processing overhead (\ie copying + DPU processing).
For example, for $R=1$ (the highest degree of redundancy), the naive copy time to DPUs is on average $31.9\times$ larger compared to the DPU processing time.
However, with CAC, this copy time is only $4.6\times$ larger than the DPU processing time on average for $R=1$.
Overall, when considering the end-to-end performance, CAC yields up to $9.5\times$ speedup when compared to the naive copy approach. 

Still, when considering the highest degree of redundancy ($R=1$), we observe that the multi-threaded CPU-based workload is slightly better ($\approx1.16\times$ faster) compared to the PIM-based workload with CAC.
Although this may suggest that the CPU's performance is better than the PIM system, it is important to note that our experiments use only 256 DPUs, which is a small fraction of the total 2560 DPUs available on a fully populated PIM server.
In contrast, 32 threads is the highest degree of parallelism the CPU-based system can attain.
As such, a full-scale PIM system equipped with CAC is therefore very likely to outperform the CPU-based system on large and highly redundant workloads.

\observ{
While data transfer remains the main bottleneck in PIM-based workloads, CAC provides huge potential to overcome this barrier, making PIM-based computation more practical and impactful.
}

\subsection{Memory overhead}
\label{eval:mem}
While techniques like CAC eliminate redundant content to gain data transfer improvements, they inevitably incur some memory overhead for index management.
Here, we do a thorough analysis of the memory overhead introduced by CAC and VByte compression.

\subpoint{In-memory index metadata.}
For each DPU, CAC maintains a hash table for storing block fingerprints and their corresponding BRB offsets. 
For each $1$KiB block of data to be copied to DPUs, we store an $8$B fingerprint and $4$B BRB offset, thus $12$B of indexing metadata per data block.
Thus, for every $N$ bytes of data to be copied to DPUs, we store $\frac{12*N}{1024}$ bytes of additional information.
This is equivalent to $12$KiB of in-memory indexing metadata for every $1$MiB of data transferred to DPUs.
We equally maintain a buffer to track the top of the BRB for each DPUs.
Each entry in this buffer is a $4$B integer, meaning for $D$ DPUs, we store $4*D$ bytes of data in memory to track the position of the BRBs.
Thus, for $N$ bytes of data copied to DPUs and $D$ DPUs, the total size of in-memory metadata (stored all through the PIM-program's lifetime) with CAC is: $\frac{12*N}{1024} + 4*D$ bytes.
If a FIFO BRB replacement technique is used, this requires a reverse lookup table to track the block fingerprints at all BRB offsets in the hash tables.
This equally requires $\frac{12*N}{1024}$ bytes of host DRAM memory.

\subpoint{DPU block offsets in \tob.}
For each $1$KiB ($1024$B) block of data fed to the DRM (which will either be copied to DPUs or not), we have to copy a $4$B offset (tracked in \tob, see \S\ref{sec:arch}) to DPU memory.
Assuming an original $N$ bytes input buffer has been split equally among $D$ DPUs, with $N_D$ bytes of data per DPU (thus $N_D = \frac{N}{D}$), we need to transfer $\frac{4*N_D}{1024} = \frac{N_D}{256}$ bytes worth of BRB offsets (size of the \tob per DPU) to each DPU.
This is equivalent to $4$KiB worth of BRB offsets per MiB of data.
If the deduplication percentage is $d\%$ for all $N$ bytes of data fed to the DRM, then $N - \frac{d*N}{100}$ bytes of input data are actually copied to the DPUs.
This value equally represents the sum of all per DPU temporary block buffers (\tbb).
The total amount of data sent to PIM memory is thus: \(N - \frac{d*N}{100} + \frac{4*D*N_D}{1024} = N(1 -\frac{d}{100} + \frac{1}{256})\) bytes.

We note that after each copy operation, \tbb and \tob are freed, thus leaving memory for other processes.

\subpoint{Compression.}
Using a compression technique like VByte requires MRAM to accommodate the decompressed buffer, after which the memory occupied by the compressed buffer can be freed.
For an original input buffer of size $N$ bytes, if the compression ratio is $r$, then we have a final compressed buffer of size $\frac{N}{r}$.
Assuming we have $D$ DPUs and thus $N_D = \frac{N}{D}$ bytes of the original data per DPU, for decompression to complete successfully, each DPU should contain sufficient MRAM memory to accommodate $N_D + \frac{N_D}{r} = N_D*(\frac{r+1}{r})$ bytes of data. 
Once decompression terminates, the $\frac{N_D}{r}$ bytes occupied by the compressed chunk can be freed for other PIM operations.

\subsection{Discussion}
\label{discussion}
\subpoint{Serial vs parallel transfers.}
All our experiments use UPMEM's serial data transfer API, \code{dpu\_copy\_to} to transfer data to DPU memory.
As discussed in \S\ref{sec:background}, the UPMEM toolchain also enables parallel transfers to multiple MRAM banks using the API \code{dpu\_push\_xfer}, provided the data sizes and MRAM destination addresses/offsets are the same.
In practice, many workloads do not satisfy these strict alignment and size requirements, which significantly limits the applicability of \code{dpu\_push\_xfer}.
For example, when using a content-aware copy approach, the size of deduplicated data per DPU and the MRAM/BRB destination addresses for data to be copied are never the same for all DPUs, and so UPMEM's parallel transfer API simply fails 100\% of the time.
As such, to have a fair evaluation, all systems used the serial data transfer API which does not impose these restrictions. 
Nevertheless, data transfer to DPUs with CAC could be parallelized without the same restrictions by allowing each DRM thread to push its deduplicated buffer once complete.
So overall, a parallelized CAC approach is expected to provide better copy performance for high-redundancy workloads when compared to UPMEM's equivalent parallel transfer API.

\textsc{pim-cache} could be integrated into UPMEM's programming toolchain, giving PIM programmers an opportunity to achieve better performance across high-redundancy workloads, without needing to develop ad-hoc data reduction techniques.
For workloads containing low redundancy, UPMEM's parallel transfer API should be leveraged if the data sizes and MRAM destination offsets are the same. 
In the following, we discuss how to fall back to UPMEM's copy API in case of low redundancy.

\subpoint{Fallback copy approach.}
Given that \textsc{pim-cache} provides benefit only for workloads with high spatial or temporal redundancy, a fallback mechanism can be integrated to handle low-redundancy situations.
This fallback approach will involve using UPMEM's default transfer API (serial or parallel) to perform the copying to DPUs. 
Implementation-wise, this could be done by maintaining a temporal redundancy score (\eg deduplication percentage) across several copy iterations and measuring the average.
If this average falls below a predefined threshold, $\tau$ (\eg $30\%$), then all subsequent transfers bypass \textsc{pim-cache} and use UPMEM's default copy approach.
We defer a full exploration of this adaptive policy to future work.

\subpoint{DPU-CPU data transfers.}
The current design of \textsc{pim-cache} focuses solely on CPU-DPU transfers, as our experience indicates that this typically accounts for most of the data transfer overhead.
However, DPU-CPU transfers equally introduce large overhead, particularly in workloads requiring inter-DPU data exchange~\cite{gilbert24}, which must go through the CPU as discussed in \S\ref{sec:background}.
While the host-side block indexing logic could be implemented DPU-side, this would likely strain already limited DPU memory which is capped at 64MiB of MRAM and 64KiB of WRAM; these should be prioritized for PIM processing.
Moreover, the degree of redundancy in DPU intermediate results may not be sufficient to justify deduplication.
Nevertheless, lightweight VByte compression could provide reasonable performance improvements in this context.
Extending \textsc{pim-cache} in this direction would lead to a more holistic solution for addressing overall data transfer overhead in UPMEM-PIM.
We equally defer this to future work.

\section{Related work}
\label{sec:rw}
In this section, we explore related work under the following categories:
\emph{(i)}~PIM-based solutions for big data processing,
\emph{(ii)}~Data reduction with deduplication and compression, and 
\emph{(iii)}~Approaches for mitigating PIM data transfer overhead.

\subpoint{PIM-based big data processing.}
Several studies have leveraged recent PIM technologies like UPMEM's to accelerate big data processing workloads.
In the area of bioinformatics, particularly genome analysis, \cite{diab2023framework} leverages UPMEM-PIM for high-throughput genome sequence alignment.
They report a $28.14\times$ speedup relative to a CPU-based approach when data transfer times between the CPU and DPU are not taken into account, and a significantly lower speedup of $2.56\times$ when the data transfer costs are taken into account.
This highlights the issue our work addresses.
Similarly, other studies like \cite{alser2020accelerating, Hur2024AcceleratingDR, lavenier16} investigate PIM as a solution to mitigate the von Neumann bottleneck with highly data-intensive genomic processing.
\cite{gomez22} conducts an extensive experimental analysis of UPMEM-PIM workloads and provide an open-source PIM benchmarking suite.
Their results equally highlight the high CPU-DPU transfer overhead introduced in various workloads.
SparseP~\cite{sparsep} proposes a PIM-based library for efficient sparse matrix-vector multiplication (SpMV), with support for various compressed matrix formats.
In addition, they provide suggestions and recommendations to PIM hardware designers on mitigating CPU-DPU data transfers, varying from ad-hoc to more generic solutions.

Our work clearly identifies the CPU-DPU data transfer problem and proposes a systematic approach which spans multiple workloads and applications, providing improved data transfer performance when data redundancy is high.

\subpoint{Data reduction with deduplication and compression.}
A large body of research has explored both inline~\cite{idedup, chunkstash, cache-dedup, light-dedup} and post-process~\cite{primary-dedup, an2013offline, meyer2012study, koller20} deduplication techniques for data reduction in storage systems.
\cite{zipline} leverages deduplication and compression techniques to improve network throughput with programmable switches.
In the area of genomics, GenoDedup~\cite{cogo21} leverages similarity-based deduplication and specialized delta-encoding for reducing genome sequencing data.
Other studies like \cite{bartus18,zhang2023deduplication} equally leverage deduplication to reduce storage costs for genomic data.

Our work builds on the theoretical foundations provided by these earlier research efforts on deduplication and data compression to provide a pragmatic solution that mitigates data transfer overhead for PIM architectures like UPMEM.

\subpoint{Approaches for mitigating PIM data transfer overhead.}
While several recent studies~\cite{gilbert24,lee2024, mpoki24, friesel23, hyun24} have explicitly acknowledged the CPU-DPU data transfer issue inherent in PIM architectures like UPMEM-PIM, very few have proposed systematic solutions to the problem.
\cite{gilbert24} conducts an extensive evaluation of various workloads in PIM, and clearly identifies how inter-DPU data transfers (which must go through the CPU) in UPMEM-PIM limits the scalability of PIM processing with current architectures. 
The authors propose a hardware solution via a PIM interconnect that provides direct communication between PIM modules.
PIM-MMU~\cite{pim-mmu} proposes a hardware/software co-design that enables efficient data transfers between DRAM and PIM memory.
 
These techniques are promising and aren't limited to high-redundancy workloads like \textsc{pim-cache}.
Nevertheless, the hardware modifications required make them infeasible in the near future.
We provide an open-source and entirely software-based solution, readily available to any PIM programmer.

\section{Conclusion}
\label{sec:conclusion}
This paper introduced a content-aware copy approach (CAC) for PIM-based architectures that require explicit data transfers between the CPU's address space and DPU memory. 
The CAC approach is premised on the idea of high data similarity (\ie spatial and temporal redundancies) in certain PIM-based workloads and leverages deduplication and compression techniques to eliminate redundant data transfers operations to DPUs.
Using both synthetic datasets, as well as real-world genome sequences, we demonstrate the effectiveness of this technique in reducing CPU-DPU data transfer overhead in UPMEM's PIM architecture.

\subpoint{Future work.}
We plan to integrate \sys into a full PIM pipeline, \eg with machine learning, and evaluate the resulting end-to-end overhead reductions. 
Additionally, we aim to evaluate its applicability to other accelerators like GPUs, which handle similar workloads and may also benefit from such data transfer optimizations.  
We also plan to integrate a fallback mechanism to dynamically determine if a workload has sufficient spatial or temporary redundancies to use \sys or simply fallback to a content-agnostic copy approach.

%

	\bibliographystyle{ACM-Reference-Format}
	\bibliography{refs}

@inproceedings{zhang-interleaving,
	author = {Zhang, Zhao and Zhu, Zhichun and Zhang, Xiaodong},
	title = {A permutation-based page interleaving scheme to reduce row-buffer conflicts and exploit data locality},
	year = {2000},
	isbn = {1581131968},
	publisher = {Association for Computing Machinery},
	address = {New York, NY, USA},
	url = {https://doi.org/10.1145/360128.360134},
	doi = {10.1145/360128.360134},
	booktitle = {Proceedings of the 33rd Annual ACM/IEEE International Symposium on Microarchitecture},
	pages = {32–41},
	numpages = {10},
	location = {Monterey, California, USA},
	series = {MICRO 33}
}

@article{im-pir,
	title={IM-PIR: In-Memory Private Information Retrieval},
	author={Mwaisela, Mpoki and Yuhala, Peterson and Felber, Pascal and Schiavoni, Valerio},
	journal={arXiv preprint arXiv:2509.06514},
	year={2025}
}

@inproceedings {jeon19,
	author = {Myeongjae Jeon and Shivaram Venkataraman and Amar Phanishayee and Junjie Qian and Wencong Xiao and Fan Yang},
	title = {Analysis of {Large-Scale} {Multi-Tenant} {GPU} Clusters for {DNN} Training Workloads},
	booktitle = {2019 USENIX Annual Technical Conference (USENIX ATC 19)},
	year = {2019},
	isbn = {978-1-939133-03-8},
	address = {Renton, WA},
	pages = {947--960},
	url = {https://www.usenix.org/conference/atc19/presentation/jeon},
	publisher = {USENIX Association},
	month = jul
}

@inproceedings {tegra21,
	author = {Anand Padmanabha Iyer and Qifan Pu and Kishan Patel and Joseph E. Gonzalez and Ion Stoica},
	title = {{TEGRA}: Efficient {Ad-Hoc} Analytics on Evolving Graphs},
	booktitle = {18th USENIX Symposium on Networked Systems Design and Implementation (NSDI 21)},
	year = {2021},
	isbn = {978-1-939133-21-2},
	pages = {337--355},
	url = {https://www.usenix.org/conference/nsdi21/presentation/iyer},
	publisher = {USENIX Association},
	month = apr
}

@inproceedings{basak21,
	author = {Basak, Abanti and Qu, Zheng and Lin, Jilan and Alameldeen, Alaa R. and Chishti, Zeshan and Ding, Yufei and Xie, Yuan},
	title = {Improving Streaming Graph Processing Performance using Input Knowledge},
	year = {2021},
	isbn = {9781450385572},
	publisher = {Association for Computing Machinery},
	address = {New York, NY, USA},
	url = {https://doi.org/10.1145/3466752.3480096},
	doi = {10.1145/3466752.3480096},	
	booktitle = {MICRO-54: 54th Annual IEEE/ACM International Symposium on Microarchitecture},
	pages = {1036–1050},
	numpages = {15},
	location = {Virtual Event, Greece},
	series = {MICRO '21}
}

@inproceedings{common-graph23,
	author = {Afarin, Mahbod and Gao, Chao and Rahman, Shafiur and Abu-Ghazaleh, Nael and Gupta, Rajiv},
	title = {CommonGraph: Graph Analytics on Evolving Data},
	year = {2023},
	isbn = {9781450399166},
	publisher = {Association for Computing Machinery},
	address = {New York, NY, USA},
	url = {https://doi.org/10.1145/3575693.3575713},
	doi = {10.1145/3575693.3575713},	
	booktitle = {Proceedings of the 28th ACM International Conference on Architectural Support for Programming Languages and Operating Systems, Volume 2},
	pages = {133–145},
	numpages = {13},
	location = {Vancouver, BC, Canada},
	series = {ASPLOS 2023}
}

@inproceedings{tgopt23,
	author = {Wang, Yufeng and Mendis, Charith},
	title = {TGOpt: Redundancy-Aware Optimizations for Temporal Graph Attention Networks},
	year = {2023},
	isbn = {9798400700156},
	publisher = {Association for Computing Machinery},
	address = {New York, NY, USA},
	url = {https://doi.org/10.1145/3572848.3577490},
	doi = {10.1145/3572848.3577490},
	booktitle = {Proceedings of the 28th ACM SIGPLAN Annual Symposium on Principles and Practice of Parallel Programming},
	pages = {354–368},
	numpages = {15},
	location = {Montreal, QC, Canada},
	series = {PPoPP '23}
}

@article{gomez-ml-pim,
	title={An experimental evaluation of machine learning training on a real processing-in-memory system},
	author={G{\'o}mez-Luna, Juan and Guo, Yuxin and Brocard, Sylvan and Legriel, Julien and Cimadomo, Remy and Oliveira, Geraldo F and Singh, Gagandeep and Mutlu, Onur},
	journal={arXiv preprint arXiv:2207.07886},
	year={2022}
}

@article{roy2021pim,
	title={PIM-DRAM: Accelerating machine learning workloads using processing in commodity DRAM},
	author={Roy, Sourjya and Ali, Mustafa and Raghunathan, Anand},
	journal={IEEE Journal on Emerging and Selected Topics in Circuits and Systems},
	volume={11},
	number={4},
	pages={701--710},
	year={2021},
	publisher={IEEE}
}

@INPROCEEDINGS{hpc-log-analytics,
	author={Park, Byung H. and Hukerikar, Saurabh and Adamson, Ryan and Engelmann, Christian},
	booktitle={2017 IEEE International Conference on Cluster Computing (CLUSTER)}, 
	title={Big Data Meets HPC Log Analytics: Scalable Approach to Understanding Systems at Extreme Scale}, 
	year={2017},
	volume={},
	number={},
	pages={758-765},
	doi={10.1109/CLUSTER.2017.113}}

@misc{ucsc,
	author       = {University of California Santa Cruz},
	title        = {{UCSC Genome Browser Home}},
	howpublished = {\url{https://hgdownload.soe.ucsc.edu/downloads.html}},
	year         = {2025},
	note         = {Accessed on 24-02-2025}
}

@misc{fasta,
	author       = {National Library of Medicine},
	title        = {{FASTA Format for Nucleotide Sequences}},
	howpublished = {\url{https://www.ncbi.nlm.nih.gov/genbank/fastaformat/}},
	year         = {2025},
	note         = {Accessed on 24-02-2025}
}

@article{diab2023framework,
	title={A framework for high-throughput sequence alignment using real processing-in-memory systems},
	author={Diab, Safaa and Nassereldine, Amir and Alser, Mohammed and G{\'o}mez Luna, Juan and Mutlu, Onur and El Hajj, Izzat},
	journal={Bioinformatics},
	volume={39},
	number={5},
	pages={btad155},
	year={2023},
	publisher={Oxford University Press}
}

@article{perez2025ucsc,
	title={The UCSC Genome Browser database: 2025 update},
	author={Perez, Gerardo and Barber, Galt P and Benet-Pages, Anna and Casper, Jonathan and Clawson, Hiram and Diekhans, Mark and Fischer, Clay and Gonzalez, Jairo Navarro and Hinrichs, Angie S and Lee, Christopher M and others},
	journal={Nucleic Acids Research},
	volume={53},
	number={D1},
	pages={D1243--D1249},
	year={2025},
	publisher={Oxford University Press}
}

@misc{upmem-mem-man,
	author       = {UPMEM},
	title        = {{UPMEM SDK}},
	howpublished = {\url{https://sdk.upmem.com/2025.1.0/031_DPURuntimeService_Memory.html}},
	year         = {2025},
	note         = {Accessed on 24-02-2025}
}

@misc{xxhash,
	author       = {XXHash},
	title        = {{XXHash}},
	howpublished = {\url{https://xxhash.com/}},
	year         = {2025},
	note         = {Accessed on 20-01-2025}
}

@misc{murmur,
	author       = {Austin Appleby},
	title        = {{XXHash}},
	howpublished = {\url{https://github.com/aappleby/smhasher}},
	year         = {2025},
	note         = {Accessed on 20-01-2025}
}

@misc{farmhash,
	author       = {Google},
	title        = {{FarmHash}},
	howpublished = {\url{https://github.com/google/farmhash/}},
	year         = {2025},
	note         = {Accessed on 20-01-2025}
}

@misc{heiser-crimes,
	author       = {Gernot Heiser},
	title        = {{Systems Benchmarking Crimes}},
	howpublished = {\url{https://gernot-heiser.org/benchmarking-crimes.html}},
	year         = {2025},
	note         = {Accessed on 10-02-2025}
}

@article{meyer2012study,
	title={A study of practical deduplication},
	author={Meyer, Dutch T and Bolosky, William J},
	journal={ACM Transactions on Storage (ToS)},
	volume={7},
	number={4},
	pages={1--20},
	year={2012},
	publisher={ACM New York, NY, USA}
}

@article{sparsep,
	author = {Giannoula, Christina and Fernandez, Ivan and Luna, Juan G\'{o}mez and Koziris, Nectarios and Goumas, Georgios and Mutlu, Onur},
	title = {SparseP: Towards Efficient Sparse Matrix Vector Multiplication on Real Processing-In-Memory Architectures},
	year = {2022},
	issue_date = {March 2022},
	publisher = {Association for Computing Machinery},
	address = {New York, NY, USA},
	volume = {6},
	number = {1},
	url = {https://doi.org/10.1145/3508041},
	doi = {10.1145/3508041},
	journal = {Proc. ACM Meas. Anal. Comput. Syst.},
	month = feb,
	articleno = {21},
	numpages = {49}
}

@inproceedings {nider20,
	author = {Joel Nider and Craig Mustard and Andrada Zoltan and Alexandra Fedorova},
	title = {Processing in Storage Class Memory},
	booktitle = {12th USENIX Workshop on Hot Topics in Storage and File Systems (HotStorage 20)},
	year = {2020},
	url = {https://www.usenix.org/conference/hotstorage20/presentation/nider},
	publisher = {USENIX Association},
	month = jul
}

@inproceedings{agrawal17,
	author = {Agrawal, Sandeep R and Idicula, Sam and Raghavan, Arun and Vlachos, Evangelos and Govindaraju, Venkatraman and Varadarajan, Venkatanathan and Balkesen, Cagri and Giannikis, Georgios and Roth, Charlie and Agarwal, Nipun and Sedlar, Eric},
	title = {A many-core architecture for in-memory data processing},
	year = {2017},
	isbn = {9781450349529},
	publisher = {Association for Computing Machinery},
	address = {New York, NY, USA},
	url = {https://doi.org/10.1145/3123939.3123985},
	doi = {10.1145/3123939.3123985},
	booktitle = {Proceedings of the 50th Annual IEEE/ACM International Symposium on Microarchitecture},
	pages = {245–258},
	numpages = {14},
	location = {Cambridge, Massachusetts},
	series = {MICRO-50 '17}
}

@INPROCEEDINGS{gao16,
	author={Gao, Mingyu and Kozyrakis, Christos},
	booktitle={2016 IEEE International Symposium on High Performance Computer Architecture (HPCA)}, 
	title={HRL: Efficient and flexible reconfigurable logic for near-data processing}, 
	year={2016},
	volume={},
	number={},
	pages={126-137},
	doi={10.1109/HPCA.2016.7446059}}

@INPROCEEDINGS{ahn15,
	author={Ahn, Junwhan and Hong, Sungpack and Yoo, Sungjoo and Mutlu, Onur and Choi, Kiyoung},
	booktitle={2015 ACM/IEEE 42nd Annual International Symposium on Computer Architecture (ISCA)}, 
	title={A scalable processing-in-memory accelerator for parallel graph processing}, 
	year={2015},
	volume={},
	number={},
	pages={105-117},
	doi={10.1145/2749469.2750386}}

@inproceedings{bailey91,
	author = {Bailey, D. H. and Barszcz, E. and Barton, J. T. and Browning, D. S. and Carter, R. L. and Dagum, L. and Fatoohi, R. A. and Frederickson, P. O. and Lasinski, T. A. and Schreiber, R. S. and Simon, H. D. and Venkatakrishnan, V. and Weeratunga, S. K.},
	title = {The NAS parallel benchmarks—summary and preliminary results},
	year = {1991},
	isbn = {0897914597},
	publisher = {Association for Computing Machinery},
	address = {New York, NY, USA},
	url = {https://doi.org/10.1145/125826.125925},
	doi = {10.1145/125826.125925},
	booktitle = {Proceedings of the 1991 ACM/IEEE Conference on Supercomputing},
	pages = {158–165},
	numpages = {8},
	location = {Albuquerque, New Mexico, USA},
	series = {Supercomputing '91}
}

@article{vbyte,
	title={Stream VByte: Faster byte-oriented integer compression},
	author={Lemire, Daniel and Kurz, Nathan and Rupp, Christoph},
	journal={Information Processing Letters},
	volume={130},
	pages={1--6},
	year={2018},
	publisher={Elsevier}
}

@inproceedings{stepanov11,
	author = {Stepanov, Alexander A. and Gangolli, Anil R. and Rose, Daniel E. and Ernst, Ryan J. and Oberoi, Paramjit S.},
	title = {SIMD-based decoding of posting lists},
	year = {2011},
	isbn = {9781450307178},
	publisher = {Association for Computing Machinery},
	address = {New York, NY, USA},
	url = {https://doi.org/10.1145/2063576.2063627},
	doi = {10.1145/2063576.2063627},
	booktitle = {Proceedings of the 20th ACM International Conference on Information and Knowledge Management},
	pages = {317–326},
	numpages = {10},
	location = {Glasgow, Scotland, UK},
	series = {CIKM '11}
}

@inproceedings{dean2009,
	author = {Dean, Jeffrey},
	title = {Challenges in building large-scale information retrieval systems: invited talk},
	year = {2009},
	isbn = {9781605583907},
	publisher = {Association for Computing Machinery},
	address = {New York, NY, USA},
	url = {https://doi.org/10.1145/1498759.1498761},
	doi = {10.1145/1498759.1498761},
	booktitle = {Proceedings of the Second ACM International Conference on Web Search and Data Mining},
	pages = {1},
	numpages = {1},
	location = {Barcelona, Spain},
	series = {WSDM '09}
}

@article{brandon2009data,
	title={Data structures and compression algorithms for genomic sequence data},
	author={Brandon, Marty C and Wallace, Douglas C and Baldi, Pierre},
	journal={Bioinformatics},
	volume={25},
	number={14},
	pages={1731--1738},
	year={2009},
	publisher={Oxford University Press}
}

@inproceedings {idedup,
	title = {{iDedup}: Latency-aware, Inline Data Deduplication for Primary Storage},
	booktitle = {10th USENIX Conference on File and Storage Technologies (FAST 12)},
	year = {2012},
	address = {San Jose, CA},
	url = {https://www.usenix.org/conference/fast12/idedup-latency-aware-inline-data-deduplication-primary-storage},
	publisher = {USENIX Association},
	month = feb
}

@inproceedings {chunkstash,
	author = {Biplob Debnath and Sudipta Sengupta and Jin Li},
	title = {{ChunkStash}: Speeding Up Inline Storage Deduplication Using Flash Memory},
	booktitle = {2010 USENIX Annual Technical Conference (USENIX ATC 10)},
	year = {2010},
	url = {https://www.usenix.org/conference/usenix-atc-10/chunkstash-speeding-inline-storage-deduplication-using-flash-memory},
	publisher = {USENIX Association},
	month = jun
}

@inproceedings {koller20,
	author = {Ricardo Koller and Raju Rangaswami},
	title = {{I/O} A High Performance Deduplication Engine with Mixed Pages},
	booktitle = {8th USENIX Conference on File and Storage Technologies (FAST 10)},
	year = {2010},
	address = {San Jose, CA },
	url = {https://www.usenix.org/conference/fast-10/io-deduplication-utilizing-content-similarity-improve-io-performance},
	publisher = {USENIX Association},
	month = feb
}

@inproceedings {primary-dedup,
	author = {Ahmed El-Shimi and Ran Kalach and Ankit Kumar and Adi Ottean and Jin Li and Sudipta Sengupta},
	title = {Primary Data {Deduplication{\textemdash}Large} Scale Study and System Design},
	booktitle = {2012 USENIX Annual Technical Conference (USENIX ATC 12)},
	year = {2012},
	isbn = {978-931971-93-5},
	address = {Boston, MA},
	pages = {285--296},
	url = {https://www.usenix.org/conference/atc12/technical-sessions/presentation/el-shimi},
	publisher = {USENIX Association},
	month = jun
}

@inproceedings {austere-cache,
	author = {Qiuping Wang and Jinhong Li and Wen Xia and Erik Kruus and Biplob Debnath and Patrick P. C. Lee},
	title = {Austere Flash Caching with Deduplication and Compression},
	booktitle = {2020 USENIX Annual Technical Conference (USENIX ATC 20)},
	year = {2020},
	isbn = {978-1-939133-14-4},
	pages = {713--726},
	url = {https://www.usenix.org/conference/atc20/presentation/wang-qiuping},
	publisher = {USENIX Association},
	month = jul
}

@article{mutlu2019processing,
	title={Processing data where it makes sense: Enabling in-memory computation},
	author={Mutlu, Onur and Ghose, Saugata and G{\'o}mez-Luna, Juan and Ausavarungnirun, Rachata},
	journal={Microprocessors and Microsystems},
	volume={67},
	pages={28--41},
	year={2019},
	publisher={Elsevier}
}

@INPROCEEDINGS{rowclone23,
	author={Seshadri, Vivek and Kim, Yoongu and Fallin, Chris and Lee, Donghyuk and Ausavarungnirun, Rachata and Pekhimenko, Gennady and Luo, Yixin and Mutlu, Onur and Gibbons, Phillip B. and Kozuch, Michael A. and Mowry, Todd C.},
	booktitle={2013 46th Annual IEEE/ACM International Symposium on Microarchitecture (MICRO)}, 
	title={RowClone: Fast and energy-efficient in-DRAM bulk data copy and initialization}, 
	year={2013},
	volume={},
	number={},
	pages={185-197},
	doi={}}

@article{alser2020accelerating,
	title={Accelerating genome analysis: A primer on an ongoing journey},
	author={Alser, Mohammed and Bing{\"o}l, Z{\"u}lal and Cali, Damla Senol and Kim, Jeremie and Ghose, Saugata and Alkan, Can and Mutlu, Onur},
	journal={IEEE Micro},
	volume={40},
	number={5},
	pages={65--75},
	year={2020},
	publisher={IEEE}
}

@inproceedings {light-dedup,
	author = {Jiansheng Qiu and Yanqi Pan and Wen Xia and Xiaojia Huang and Wenjun Wu and Xiangyu Zou and Shiyi Li and Yu Hua},
	title = {{Light-Dedup}: A Light-weight Inline Deduplication Framework for {Non-Volatile} Memory File Systems},
	booktitle = {2023 USENIX Annual Technical Conference (USENIX ATC 23)},
	year = {2023},
	isbn = {978-1-939133-35-9},
	address = {Boston, MA},
	pages = {101--116},
	url = {https://www.usenix.org/conference/atc23/presentation/qiu-jiansheng},
	publisher = {USENIX Association},
	month = jul
}

@ARTICLE{wu18,
	author={Wu, Huijun and Wang, Chen and Fu, Yinjin and Sakr, Sherif and Lu, Kai and Zhu, Liming},
	journal={IEEE Transactions on Parallel and Distributed Systems}, 
	title={A Differentiated Caching Mechanism to Enable Primary Storage Deduplication in Clouds}, 
	year={2018},
	volume={29},
	number={6},
	pages={1202-1216},
	doi={10.1109/TPDS.2018.2790946}}

@inproceedings{an2013offline,
	title={Offline deduplication-aware block separation for solid state disk},
	author={An, Jeongcheol and Shin, Dongkun},
	booktitle={11th USENIX Conference on File and Storage Technologies (FAST 13)},
	year={2013}
}

@inproceedings {cache-dedup,
	author = {Wenji Li and Gregory Jean-Baptise and Juan Riveros and Giri Narasimhan and Tony Zhang and Ming Zhao},
	title = {{CacheDedup}: In-line Deduplication for Flash Caching},
	booktitle = {14th USENIX Conference on File and Storage Technologies (FAST 16)},
	year = {2016},
	isbn = {978-1-931971-28-7},
	address = {Santa Clara, CA},
	pages = {301--314},
	url = {https://www.usenix.org/conference/fast16/technical-sessions/presentation/li-wenji},
	publisher = {USENIX Association},
	month = feb
}

@article{pimpam,
	author = {Cai, Shuangyu and Tian, Boyu and Zhang, Huanchen and Gao, Mingyu},
	title = {PimPam: Efficient Graph Pattern Matching on Real Processing-in-Memory Hardware},
	year = {2024},
	issue_date = {June 2024},
	publisher = {Association for Computing Machinery},
	address = {New York, NY, USA},
	volume = {2},
	number = {3},
	url = {https://doi.org/10.1145/3654964},
	doi = {10.1145/3654964},
	
	journal = {Proc. ACM Manag. Data},
	month = may,
	articleno = {161},
	numpages = {25}
}

@ARTICLE {lee2024,
	author = {D. Lee and B. Hyun and T. Kim and M. Rhu},
	journal = {IEEE Computer Architecture Letters},
	title = {Analysis of Data Transfer Bottlenecks in Commercial {PIM} Systems: A Study with {UPMEM-PIM}},
	year = {2024},
	volume = {},
	number = {01},
	issn = {1556-6064},
	pages = {1-4},
	doi = {10.1109/LCA.2024.3387472},
	publisher = {IEEE Computer Society},
	address = {Los Alamitos, CA, USA},
	month = {apr}
}

@techreport{upmem-pim,
	author = {{UPMEM}},
	title = {{UPMEM} Processing In-Memory ({PIM}): ultra-efficient acceleration for data-intensive applications},
	year = {2022},
	type = {White paper},
	month = aug,
}

@INPROCEEDINGS{mpoki24,
  author={Mwaisela, Mpoki and Hari, Joel and Yuhala, Peterson and Ménétrey, Jämes and Felber, Pascal and Schiavoni, Valerio},
  booktitle={2024 43rd International Symposium on Reliable Distributed Systems (SRDS)}, 
  title={Evaluating the Potential of In-Memory Processing to Accelerate Homomorphic Encryption: Practical Experience Report}, 
  year={2024},
  volume={},
  number={},
  pages={92-103},
  doi={10.1109/SRDS64841.2024.00019}}

@inproceedings{nider21,
	author    = {Joel Nider and Craig Mustard and Andrada Zoltan and John Ramsden and Larry Liu and Jacob Grossbard and Mohammad Dashti and Romaric Jodin and Alexandre Ghiti and Jordi Chauzi and Alexandra Fedorova},
	title     = {A Case Study of Processing-in-Memory in off-the-Shelf Systems},
	booktitle = {2021 USENIX Annual Technical Conference (USENIX ATC 21)},
	year      = {2021},
	isbn      = {978-1-939133-23-6},
	pages     = {117--130},
	url       = {https://www.usenix.org/conference/atc21/presentation/nider},
	publisher = {USENIX Association},
	month     = jul
}

@article{gilbert24,
	author     = {Jonatan, Gilbert and Cho, Haeyoon and Son, Hyojun and Wu, Xiangyu and Livesay, Neal and Mora, Evelio and Shivdikar, Kaustubh and Abell\'{a}n, Jos\'{e} L. and Joshi, Ajay and Kaeli, David and Kim, John},
	title      = {Scalability Limitations of Processing-in-Memory using Real System Evaluations},
	year       = {2024},
	issue_date = {March 2024},
	publisher  = {Association for Computing Machinery},
	address    = {New York, NY, USA},
	volume     = {8},
	number     = {1},
	url        = {https://doi.org/10.1145/3639046},
	doi        = {10.1145/3639046},
	journal    = {Proc. ACM Meas. Anal. Comput. Syst.},
	month      = {feb},
	articleno  = {5},
	numpages   = {28}
}

@article{gomez22,
	author   = {Gómez-Luna, Juan and Hajj, Izzat El and Fernandez, Ivan and Giannoula, Christina and Oliveira, Geraldo F. and Mutlu, Onur},
	journal  = {IEEE Access},
	title    = {Benchmarking a New Paradigm: Experimental Analysis and Characterization of a Real Processing-in-Memory System},
	year     = {2022},
	volume   = {10},
	number   = {},
	pages    = {52565-52608},
	doi      = {10.1109/ACCESS.2022.3174101}
}

@INPROCEEDINGS{hyun24,
	author={Hyun, Bongjoon and Kim, Taehun and Lee, Dongjae and Rhu, Minsoo},
	booktitle={2024 IEEE International Symposium on High-Performance Computer Architecture (HPCA)}, 
	title={Pathfinding Future PIM Architectures by Demystifying a Commercial PIM Technology}, 
	year={2024},
	volume={},
	number={},
	pages={263-279},
	doi={10.1109/HPCA57654.2024.00029}}

@INPROCEEDINGS {pim-mmu,
	author = { Lee, Dongjae and Hyun, Bongjoon and Kim, Taehun and Rhu, Minsoo },
	booktitle = { 2024 57th IEEE/ACM International Symposium on Microarchitecture (MICRO) },
	title = {{ PIM-MMU: A Memory Management Unit for Accelerating Data Transfers in Commercial PIM Systems }},
	year = {2024},
	volume = {},
	ISSN = {},
	pages = {627-642},
	doi = {10.1109/MICRO61859.2024.00053},
	url = {https://doi.ieeecomputersociety.org/10.1109/MICRO61859.2024.00053},
	publisher = {IEEE Computer Society},
	address = {Los Alamitos, CA, USA},
	month =Nov}

@inproceedings{saransh21,
	author    = {Gupta, Saransh and Rosing, Tajana Šimunić},
	booktitle = {2021 58th ACM/IEEE Design Automation Conference (DAC)},
	title     = {Invited: Accelerating Fully Homomorphic Encryption with Processing in Memory},
	year      = {2021},
	volume    = {},
	number    = {},
	pages     = {1335-1338},
	doi       = {10.1109/DAC18074.2021.9586285}
}

@inproceedings{friesel23,
	author = {Friesel, Birte and L\"{u}tke Dreimann, Marcel and Spinczyk, Olaf},
	title = {A Full-System Perspective on UPMEM Performance},
	year = {2023},
	isbn = {9798400703003},
	publisher = {Association for Computing Machinery},
	address = {New York, NY, USA},
	url = {https://doi.org/10.1145/3609308.3625266},
	doi = {10.1145/3609308.3625266},
	booktitle = {Proceedings of the 1st Workshop on Disruptive Memory Systems},
	pages = {1–7},
	numpages = {7},
	location = {Koblenz, Germany},
	series = {DIMES '23}
}

@article{Hur2024AcceleratingDR,
	title={Accelerating DNA Read Mapping with Digital Processing-in-Memory},
	author={Rotem Ben Hur and Orian Leitersdorf and Ronny Ronen and Lidor Goldshmidt and Idan Magram and Lior Kaplun and Leonid Yavitz and Shahar Kvatinsky},
	journal={ArXiv},
	year={2024},
	volume={abs/2411.03832},
	url={https://api.semanticscholar.org/CorpusID:273850423}
}

@INPROCEEDINGS{lavenier16,
	author={Lavenier, Dominique and Roy, Jean-Francois and Furodet, David},
	booktitle={2016 IEEE International Conference on Bioinformatics and Biomedicine (BIBM)}, 
	title={DNA mapping using Processor-in-Memory architecture}, 
	year={2016},
	volume={},
	number={},
	pages={1429-1435},
	doi={10.1109/BIBM.2016.7822732}}

@misc{grch38,
	author       = {National Library of Medicine},
	title        = {{Genome assembly GRCh38}},
	howpublished = {\url{https://www.ncbi.nlm.nih.gov/datasets/genome/GCF_000001405.26/}},
	note         = {Accessed on 20-01-2025}
}

@article{schneider2017evaluation,
	title={Evaluation of GRCh38 and de novo haploid genome assemblies demonstrates the enduring quality of the reference assembly},
	author={Schneider, Valerie A and Graves-Lindsay, Tina and Howe, Kerstin and Bouk, Nathan and Chen, Hsiu-Chuan and Kitts, Paul A and Murphy, Terence D and Pruitt, Kim D and Thibaud-Nissen, Fran{\c{c}}oise and Albracht, Derek and others},
	journal={Genome research},
	volume={27},
	number={5},
	pages={849--864},
	year={2017},
	publisher={Cold Spring Harbor Lab}
}

@article{schbath2012mapping,
	title={Mapping reads on a genomic sequence: an algorithmic overview and a practical comparative analysis},
	author={Schbath, Sophie and Martin, V{\'e}ronique and Zytnicki, Matthias and Fayolle, Julien and Loux, Valentin and Gibrat, Jean-Fran{\c{c}}ois},
	journal={Journal of Computational Biology},
	volume={19},
	number={6},
	pages={796--813},
	year={2012},
	publisher={Mary Ann Liebert, Inc. 140 Huguenot Street, 3rd Floor New Rochelle, NY 10801 USA}
}

@article{dunham2013contemporary,
	title={Contemporary, yeast-based approaches to understanding human genetic variation},
	author={Dunham, Maitreya J and Fowler, Douglas M},
	journal={Current opinion in genetics \& development},
	volume={23},
	number={6},
	pages={658--664},
	year={2013},
	publisher={Elsevier}
}

@article{ben2024dart,
	title={DART-PIM: DNA read mApping acceleRaTor Using Processing-In-Memory},
	author={Ben-Hur, Rotem and Leitersdorf, Orian and Ronen, Ronny and Goldshmidt, Lidor and Magram, Idan and Kaplun, Lior and Yavitz, Leonid and Kvatinsky, Shahar},
	journal={arXiv preprint arXiv:2411.03832},
	year={2024}
}

@article{aganezov2022complete,
	title={A complete reference genome improves analysis of human genetic variation},
	author={Aganezov, Sergey and Yan, Stephanie M and Soto, Daniela C and Kirsche, Melanie and Zarate, Samantha and Avdeyev, Pavel and Taylor, Dylan J and Shafin, Kishwar and Shumate, Alaina and Xiao, Chunlin and others},
	journal={Science},
	volume={376},
	number={6588},
	pages={eabl3533},
	year={2022},
	publisher={American Association for the Advancement of Science}
}

@ARTICLE{cogo21,
	author={Cogo, Vinicius and Paulo, João and Bessani, Alysson},
	journal={IEEE Transactions on Computers}, 
	title={GenoDedup: Similarity-Based Deduplication and Delta-Encoding for Genome Sequencing Data}, 
	year={2021},
	volume={70},
	number={5},
	pages={669-681},
	doi={10.1109/TC.2020.2994774}}

@article{treangen2012repetitive,
	title={Repetitive DNA and next-generation sequencing: computational challenges and solutions},
	author={Treangen, Todd J and Salzberg, Steven L},
	journal={Nature Reviews Genetics},
	volume={13},
	number={1},
	pages={36--46},
	year={2012},
	publisher={Nature Publishing Group UK London}
}

@inproceedings{zipline,
	author = {Vaucher, S\'{e}bastien and Yazdani, Niloofar and Felber, Pascal and Lucani, Daniel E. and Schiavoni, Valerio},
	title = {ZipLine: in-network compression at line speed},
	year = {2020},
	isbn = {9781450379489},
	publisher = {Association for Computing Machinery},
	address = {New York, NY, USA},
	url = {https://doi.org/10.1145/3386367.3431302},
	doi = {10.1145/3386367.3431302},
	booktitle = {Proceedings of the 16th International Conference on Emerging Networking EXperiments and Technologies},
	pages = {399–405},
	numpages = {7},
	location = {Barcelona, Spain},
	series = {CoNEXT '20}
}

@article{zhang2023deduplication,
	title={Deduplication improves cost-efficiency and yields of de novo assembly and binning of shotgun metagenomes in microbiome research},
	author={Zhang, Zhiguo and Zhang, Lu and Zhang, Guoqing and Zhao, Ze and Wang, Hui and Ju, Feng},
	journal={Microbiology Spectrum},
	volume={11},
	number={2},
	pages={e04282--22},
	year={2023},
	publisher={American Society for Microbiology 1752 N St., NW, Washington, DC}
}

@INPROCEEDINGS{bartus18,
	author={Bartus, Paul and Arzuaga, Emmanuel},
	booktitle={2018 IEEE International Congress on Big Data (BigData Congress)}, 
	title={GDedup: Distributed File System Level Deduplication for Genomic Big Data}, 
	year={2018},
	volume={},
	number={},
	pages={120-127},
	doi={10.1109/BigDataCongress.2018.00023}}

\end{document}